\documentclass{llncs}
\usepackage{makeidx}  
\usepackage{enumitem}
\usepackage{color}
\usepackage{amsfonts,amsmath}
\usepackage{url}
\usepackage{hyperref}
\usepackage{graphicx}
\usepackage{caption}
\usepackage{subcaption}
\usepackage{eurosym}
\usepackage{wrapfig}
\title{Determining the veracity of rumours on Twitter}
\author{Georgios Giasemidis\inst{1} \and Colin Singleton\inst{1} 
	\and Ioannis Agrafiotis\inst{2} \and Jason R. C. Nurse\inst{2} 
	\and Alan Pilgrim\inst{3}, Chris Willis\inst{3} and Danica Vukadinovic Greetham\inst{4}}
\institute{
	CountingLab LTD, Reading, U.K., \email{\{georgios,colin\}@countinglab.co.uk}
	\and
	Department of Computer Science, University of Oxford, Oxford, U.K.,
	\email{\{ioannis.agrafiotis,jason.nurse\}@cs.ox.ac.uk}
	\and
	BAE Systems Applied Intelligence, Chelmsford, U.K.,
	\email{\{alan.pilgrim,chris.willis\}@baesystems.com}
	\and
	Department of Mathematics and Statistics, University of Reading, Reading, U.K.
	\email{d.v.greetham@reading.ac.uk}
	}

\begin{document}
	\maketitle
	
	\begin{abstract}
	While social networks can provide an ideal platform for up-to-date information from individuals across the world, it has also proved to be a place where rumours fester and accidental or deliberate misinformation often emerges. In this article, we aim to support the task of making sense from social media data, and specifically, seek to build an autonomous message-classifier that filters relevant and trustworthy information from Twitter. For our work, we collected about 100 million public tweets, including users' past tweets, from which we identified 72 rumours (41 true, 31 false). We considered over 80 trustworthiness measures including the authors' profile and past behaviour, the social network connections (graphs), and the content of tweets themselves. We ran modern machine-learning classifiers over those measures to produce trustworthiness scores at various time windows from the outbreak of the rumour. Such time-windows were key as they allowed useful insight into the progression of the rumours. From our findings, we identified that our model was significantly more accurate than similar studies in the literature. We also identified critical attributes of the data that give rise to the trustworthiness scores assigned. Finally we developed a software demonstration that provides a visual user interface to allow the user to examine the analysis.
	\end{abstract}
	
	\section{Introduction and related work}
	\label{intro}
	
Nowadays, the social media play an essential role in our everyday lives. The majority of people use social networks as their main source of information  \cite{reutersinst2015,pwerc}.  However, sources of information might be trusted, unreliable, private, invalidated or ambiguous. Rumours, for example, might be true or false and started accidentally or perhaps maliciously.
In situations of crisis, identifying rumours at an early stage in social media is crucial for decision making. The example of London Riots of 2011 is characteristic. After the events unfolded, The Guardian provided an informative graphic of the initiation and progress of a number of rumours~\cite{guardian}. This analysis showed that categorising social media information into rumours, and analysing the content and their source, may shed light into the veracity of these rumours. Moreover, it could support emergency services in obtaining a comprehensive picture in times of crisis and make better-informed decisions. 

Recently there has been progress from experts from different fields of academia in exploring the characteristics of the data source and content that will enable us to determine the veracity of information in an autonomous and efficient manner~\cite{kwon2013prominent,seo2012identifying,zubiaga2015towards}. Key concepts include information quality, which can be defined as an assessment or measure of how fit an information object is for use, and information trustworthiness, which is the likelihood that a piece of information will preserve a user's trust, or belief, in it~\cite{nurse2011information}. These concepts may overlap and indeed, increasing one (e.g., quality) may lead to an increase in the other (e.g., trustworthiness). Other relevant factors include Accuracy (Free-of-error), Reliability, Objectivity (Bias), Believability (Likelihood, Plausibility of arguments), Popularity, Competence and Provenance~\cite{wang1996beyond,mai2013quality,kelton2008trust,gil2007towards,lukyanenko2015information}.

To date, there has been a significant number of articles, in both academia and industry, that have been published on the topic of information quality and trustworthiness online, particularly in the case of Twitter. Castillo et al.~\cite{castillo2011information,castillo2013predicting} focus on developing automatic methods for assessing the credibility of posts on Twitter. They utilise a machine learning approach to the problem and for their analysis use a vast range of features grouped according to whether they are user-based, topic-based or propagation-based. Nurse et al.~\cite{nurse2014two,nurse2013supporting} have also aimed towards developing a wider framework to support the assessment of the trustworthiness of information. This framework builds on trust and quality metrics such as those already reviewed, and outlines a policy-based approach to measurement. The key aspect of this approach is that it allows organisations and users to set policies to mediate content and secondly, to weight the importance of individual trust factors (e.g., expressing that for a particular context, location is more important than corroboration). The result is a tailored trustworthiness score for information suited to the individual's unique requirements.

Having established a view on the characteristics of information quality and trustworthiness researchers focused on designing systems for rumour detection. Kwon et al.~\cite{kwon2013prominent} examined how rumours spread in social media and which characteristics may provide evidence in identifying rumours. The authors focused on three aspects of diffusion of a rumour, namely the temporal, the structural and the linguistic and identified key differences in the spread of rumours and non-rumours. Their results suggest that they were able to identify rumours with up to 92\% of accuracy. In~\cite{seo2012identifying} the authors provide an approach to identify the source of a false rumour and assess the likelihood that a specific information is true or not. They construct a directed graph where vertices denote users and edges denote information flows. They add monitor nodes reporting on data they receive and identify false rumours based on which monitoring nodes receive specific information and which do not. 

Another approach is presented in~\cite{zubiaga2015towards}, where the authors introduce a new definition for rumours and provide a novel methodology on how to collect and annotate tweets associated with an event. In contrast to other approaches which depend on predefining a set of rumours and then associating tweets to these, this methodology involves reading the replies to tweets and categorising them to stories or threads. It is a tool intended to facilitate the process of developing a machine learning approach to automatically identify rumours.

Finally, one of the most comprehensive works is presented in~\cite{vosoughi2015automatic} where various models are tested aiming for detecting and verifying rumours on Twitter. The detection of different rumours about a specific event is achieved through clustering of assertive arguments regarding a fact. By applying logistic regression on several semantic and syntactic features, the authors are able to identify with 90\% accuracy the various assertions. Regarding the verification of a rumour, the models utilise features considering the diffusion of information; feature are elicited from: linguistics, user-related aspects and temporal propagation dynamics. Hidden Markov Models are then applied to predict the veracity of the rumour; these are trained on a set or rumours whose veracity has been confirmed beforehand based on evidence from trusted websites.

Our review on the literature indicates that user and content features are found to be helpful at distinguishing trustworthy content. Moreover the temporal aspect of the aforementioned features denoting the propagation dynamics in Twitter may provide useful insights into distinguishable differences between the spread of truthful and falsified rumours. Our reflection suggests that all the approaches are using manually annotated tweets or similar datasets for the training period. The annotations denote the veracity of the rumour and indicate the event the rumour describes. Syntactic and semantic features are then extracted to aid the process of identifying events and classifying rumours regarding these events. There is not an outperforming approach and most of the models require 6-9 hours before accurately predicting the veracity. Understanding the literature on information  trustworthiness and how concepts from linguistic analysis and machine learning are applied to capture patterns of propagating information is the first and decisive step towards a system able to identify and determine the veracity of a rumour. The lessons learnt from this review are the foundations for the requirements of the system.

This paper builds on existing literature and presents a novel system which is able to collect information from social media, classify this information into rumours and determine their veracity. We collected data from Twitter, categorised these into rumours and produced a number of features for the machine learning techniques. We train and validate our dataset and compare our model to other studies in the literature. We also aim to do better than existing work and are exploring a way to visualise our various findings in a user interface.
In what follows, Section~\ref{data} reports on the data collection process and introduces  the methodology used to analyse the data, while Section~\ref{analysis} describes the analysis and model selection process. Section~\ref{results} presents the outcome of our system when applied to the collected rumours and compares our results with the results of other systems publicly available. Section~\ref{discussion} concludes the paper and identifies opportunities for future work. Finally, to keep the discussion on the modelling aspects comprehensive and compact in these sections, we present further details and by-products of our research in the Appendices.

	\section{Approach and Methodology}
	\label{data}
	We focus on messages and data from Twitter for three key reasons. First, Twitter is regarded as one of the top social networks \cite{reutersinst2015}.  
	Particularly, in emergency events, Twitter is the first choice of many for updated information, due to its continuous live feed and short length of the messages~\cite{pwerc}. 
	Second, the majority of messages on Twitter are publicly available.
	Third, Twitter's API allows us to collect the high volume of data, e.g. messages, users’ information, etc., required to build a rumour classifier. 
	
	In this study we define a rumour as~\cite{cambridge_rumour}:\textit{``An unofficial interesting story or piece of news that might be true or invented, and quickly spreads from person to person".}
	A rumour consists of all tweets from the beginning of the rumour until its verification from two trusted sources. Trusted sources are considered news agencies with global reputation, e.g. the BBC,  Reuters, the CNN, the Associated Press and a few others. For every rumour we collect four sets of data: (i) the tweets (e.g. text, timestamp, retweet/quote/reply information etc.), (ii) the users' information (e.g. user id, number of posts/followers/friends etc.), (iii) the users' followees (friends) and (iv) the users' most recent 400 tweets prior the start of the rumour, see Appendix \ref{data_collection} for a step-by-step guide on data collection. 
	In total we collected about 100 million public tweets, including users' past tweets. We  found 72 rumours, from which 41 are true and 31 are false. These rumours span diverse events, among which are: the Paris attacks in November 2015, the Brussels attacks in March 2016, the car bomb attack in Ankara in March 2016,  earthquakes in Taiwan (February 2016) and Indonesia (March 2016), train incidents near London and rumours from sports and politics, see Appendix \ref{summary_stats} for a summary statistics of these rumours. An event may contribute with more than one rumour.

	For modelling purposes we need the fraction of tweets in a rumour that support, deny or are neutral to the rumour. For this reason all the tweets are annotated as being either in favour ($+1$), against ($-1$) or neutral to ($0$) the rumour. Regarding annotation,  all retweets were assigned the same tag as their source tweet. Tweets in non-English languages that could not be easily translated were annotated as neutral. There are rumours for which this process can be automated and others which require manual tagging. 
	
	Linguistic characteristics of the messages play an important role in rumour classification \cite{castillo2013predicting,kwon2013prominent}. Extracting a text's sentiment and other linguistic characteristics can be a challenging problem which requires a lot of effort, modelling and training data. Due to the complexity of this task we used an existing well-tested tool, the Linguistic Inquiry and Word Count (LIWC), version LIWC2015 \cite{liwc2015}, which has up-to-date dictionaries consisting of about 6,400 words reflecting different emotions, thinking styles, social concerns, and even parts of speech. All collected tweets were analysed through this software. For each tweet the following attributes were extracted: (i) positive and (ii) negative sentiment score, fraction of words which represent (iii) insight, (iv) cause, (v) tentative, (vi) certainty, (vii) swearing and (viii) online abbreviations (e.g. b4 for the word before). 

	\textbf{Propagation Graph.} 
	An important set of features in the model is the propagation based set of features. All propagation features are extracted from the propagation graph or forest. A propagation forest consists of one or more connected propagation trees. A propagation tree consists of the source tweet and all of its retweets, see Appendix \ref{making_prop_graph} for further details on making a Twitter propagation tree.

	\subsection{Classifiers}
	\label{classifiers}
	
	We approach the rumour identification problem as a supervised binary classification problem, i.e. training and building a model to identify whether a rumour is true or false. This is a well-studied problem in the machine-learning field and many different methods have been developed over the years \cite{bishop2006pattern,koller2009probabilistic}. 
 
	A classifier requires a set of $N$ observations $O=\{\vec{X}_i, i=1, \ldots,N\}$ where each observation, $\vec{X}_i=\left(f_1,\ldots,f_M\right)$, is an $M$-dimensional vector of features, $f_i$. The observations are labelled into $K$ distinct classes, $\{C_i,i=1,\ldots,K\}$. Our classification problem has $N = 72$ observations (i.e. rumours) in $K=2$ classes (true or false) and $M=87$ features. The techniques that we use in this study are Logistic Regression, Support Vector Machines (SVM), with both a linear and a Radial Basis Function (RBF) kernel to investigate for both linear and non-linear relationship between the features and the classes \cite{bishop2006pattern,smola2004tutorial}, Random Forest, Decision Tree (the CART algorithm), Na\"{\i}ve Bayes and Neural Networks.
	
	We assess all models using k-fold cross-validation. The purpose of the cross-validation is to avoid an overly optimistic estimate of performance from training and testing on the same data resulting in overfitting, as the model is trained and validated on different sets of data. In this study we use $k = 10$ folds. The cross-validation method requires a classification metric to validate a model. In the literature there are several classification metrics \cite{powers2011evaluation,castillo2013predicting}, the most popular ones are: (i) accuracy, (ii) $F_1$-score, (iii) area under ROC Curve (AUC) and (iv) Cohen's kappa. We choose to cross validate the models using the $F_1$-score.
			
	Most of the statistical and machine learning methods that we use have already been developed and tested by the Python community. A popular and well-tested library is the \textit{scikit-learn}\footnote{\href{http://scikit-learn.org/}{http://scikit-learn.org/}} that we frequently use. For the neural networks classifier we used the \textit{neurolab}\footnote{\href{https://pythonhosted.org/neurolab/}{https://pythonhosted.org/neurolab/}} library.

	\subsection{Features}
	\label{feats}
	
	The rumour's features can be grouped into three broad categories, namely message-based, user-based and network-based.  
	The message-based (or linguistic) features are calculated as follows. Each tweet has a number of attributes, which can be a binary indicator, for example whether the tweet contains a URL link, or a real number, e.g. the number of words in the tweet. Having calculated all attributes we aggregate them at the rumour level. This can be, for example, the fraction of tweets in the rumour containing a URL link (for categorical attributes) or the average number of words in tweets (for continuous attributes). Both of these aggregate variables become features of the rumour. However the aggregation method can be more complicated than a fraction or an average. In particular we would like to quantify the difference in the attributes between tweets that support the rumour and those that deny it. The idea is that users who support a rumour might have different characteristics, language or behaviour, from users that deny it. To capture this difference we use an aggregation function which can be represented as:
	\begin{equation}
	\label{aggregation1}
		f_i = \frac{S^{(i)}+N^{(i)}+1}{A^{(i)}+N^{(i)}+1},
	\end{equation}
	where $f_i$ is the $i$-th feature and $S^{(i)}, N^{(i)}, A^{(i)}$ stand for support, neutral and against respectively. In mathematical terms let $D_j^{(i)}$  be the value of the $i$-th attribute of the $j$-th tweet, $j=1,\ldots,R$, $R$ being the total number of tweets in a rumour. This can be either $D_j^{(i)}\in\{0,1\}$ for binary attributes or $D_j^{(i)}\in \mathbb{R}$ for continuous ones. Also, let $B_j \in \{-1,0,1\}$ be the annotation of the $j$-th tweet. Hence we define,
	\begin{equation*}
	S^{(i)} = \frac{\sum_{j \in \{k| B_k=1\}}D^{(i)}_j}{\sum_{j \in \{k| B_k=1\}}1}, ~
	N^{(i)} = \frac{\sum_{j \in \{k| B_k=0\}}D^{(i)}_j}{\sum_{j \in \{k| B_k=0\}}1}, ~
	A^{(i)} = \frac{\sum_{j \in \{k| B_k=-1\}}D^{(i)}_j}{\sum_{j \in \{k| B_k=-1\}}1}.
	\end{equation*}

	For a practical example on the application of the above formulae, see Appendix \ref{practexample}. The aggregation formula \eqref{aggregation1} allows us to compare an attribute of the supporting tweets to an attribute of the against tweets. The neutral term in eq. \eqref{aggregation1} reduces the extremities of the ratio where there are a lot of neutral viewpoints. The unit term is a regularisation term, ensuring the denominator is always strictly positive. 
	
	There are a few attributes for which we do not follow this aggregation rule. These are the fraction of tweets that support or deny the rumour where we simply use the fractions. Additionally all the sentiment attributes can take negative values, making the denominator of eq. \eqref{aggregation1} zero or negative. For all sentiment attributes we aggregate by taking the difference between the sentiment of tweets that support the rumour and the sentiment of tweets that deny it, i.e. $S^{(i)}-A^{(i)}$. Additionally, some linguistic attributes were extracted using the popular Python library for natural language processing, \textit{nltk}\footnote{\href{http://www.nltk.org/}{http://www.nltk.org/}}.
	
	The user-based features are extracted in a similar manner focusing on the user attributes. For example, two user-attributes are whether a user has a verified account (binary) and the number of followers of a user (continuous). These attributes are aggregated to the rumour level using eq. \eqref{aggregation1}, counting each user who contributed to the rumour only once. If a user contributed with both a supporting and a refuting tweet then its attribute contributes to both the support, $S^{(i)}$, and against, $A^{(i)}$, terms.
	
	The network-based features are estimated through the propagation graph, which is constructed using the \textit{networkx}\footnote{\href{https://networkx.github.io/}{https://networkx.github.io/}} Python library. It becomes evident that three propagation graphs are required; a graph consisting of tweets that support the rumour, a graph of tweets neutral to the rumour and a graph of tweets against the rumour. From each graph we calculate a number of attributes. These network-based attributes are aggregated to the rumour feature using eq.~\eqref{aggregation1}. 
	
	
	The feature names should be treated with caution in the rest of the paper. For example, when we refer to the feature ``user’s followers" we actually mean the feature related to the user's followers through expression eqn. \eqref{aggregation1}. Exceptions are the fraction of tweets that deny the rumour, the fraction of tweets that support the rumour and all the sentiment-related features which are aggregated using expression $S^{(i)}-A^{(i)}$. 
	Therefore when we say that the feature ``user's followers" is important we don't refer to the actual number of users' followers. We hope this causes no further confusion to the reader.
	
	\textbf{Time-series Features.} One of the main goals and novel contributions of this study overall is to determine the veracity of a rumour as early as possible.  We therefore wanted to find out how quickly we could deduce the veracity. For this reason we split every rumour into 20 time-intervals and extract all the features for the subset of tweets from the start of the rumour until the end of each time-period. In this way we develop a time-series of the features which we will use to estimate the time evolution of the veracity of the rumours.

	\section{Analysis}
	\label{analysis}
	
	In this section we analyse and present the results from the cross-validation process. In particular we aim to address four key questions: (i) What is the best method for reduction of the feature space? (ii) What is the best-performing classification method? (iii) What are the optimal parameters for the selected classifiers? (iv) What are the features in the final models? 
	
	\textbf{Reduction of the Feature Space.}
	Our dataset consists of 72 observations and 87 features. The number of features is large compared to the number of observations. Using more features than necessary results in overfitting and decreased performance \cite{guyon2003a}. Feature-space reduction is a very active field of research and different techniques have been developed \cite{guyon2003a}. The importance of dimension reduction of the feature space is two-fold; first it is a powerful tool to avoid overfitting and secondly aims to reduce complexity and time of computational tasks \cite{guyon2003a,verleysen2005curse}. 
	We considered four methods of feature reduction, see Appendix \ref{feat_reduction} for further details. We apply each method to each classifier separately and assess it using k-fold cross validation. We found that the best method for reducing the feature space is a forward selection deterministic wrapper method, which we use in the rest of this study.
	
\begin{figure}
		\begin{center}
			\includegraphics[scale=0.5]{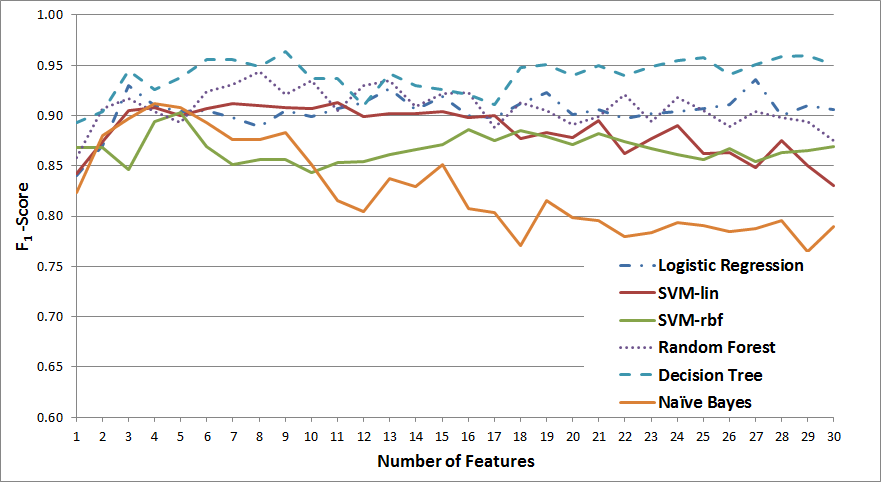}
			\caption{$F_1$-score as a function of the number of features for six classification methods. }
			\label{fig:select_classifier}
		\end{center}
	\end{figure}

	\textbf{Selecting Classifier.}
	To select the best performing classifiers, we perform a k-fold cross validation on each classifier for a number of features selected using a forward selection deterministic wrapper method. The results for \textit{scikit-learn}'s default parametrisations are plotted in Figure \ref{fig:select_classifier}.
	
	From the plot it becomes evident that, for this data set, the Decision Tree is the best performing method and the Random Forest follows. Clearly the Na\"{\i}ve Bayes and the SVM-rbf are under-performing. Logistic regression performs slightly better than the SVM-linear\footnote{We abandoned Neural Networks at early stages as the library implementation used was very slow and the results were underperforming.}. These observations are further explained and quantified in Appendix \ref{select_classifier}.
	Therefore, we select and proceed with the three-best performing methods, i.e. Logistic Regression, Random Forest, and Decision Tree. For each of the three selected classifiers we perform further analysis to optimise its input parameters to further improve its performance. 
	These parameters are the regularisation strength for Logistic Regression, the number of trees for Random Forest and the splitting criterion function for Decision Tree. We use those parameters that maximise the average $F_1$-score through cross-validation.
	 	
\begin{figure}
		\begin{center}
			\includegraphics[scale=0.5]{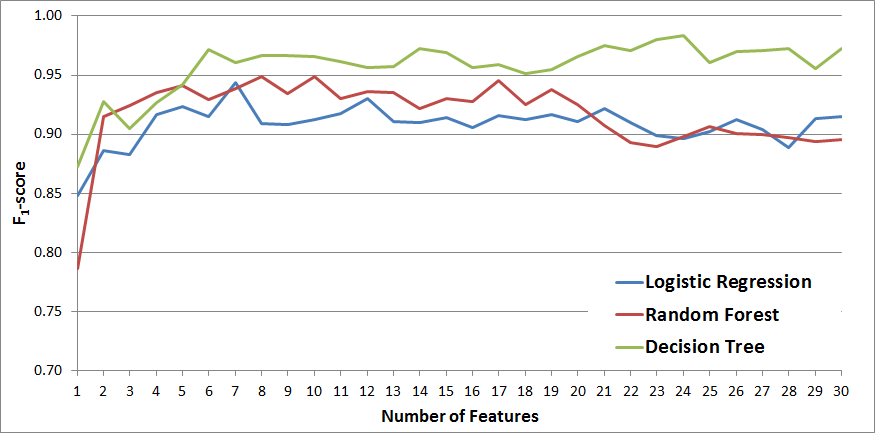}
			\caption{$F_1$-score for Logistic Regression, Random Forest and Decision Tree tuned to their optimal parameters as a function of the number of input features.}
			\label{three_classifiers}
		\end{center}
	\end{figure}

	\textbf{Best Features.}
	Having selected the best-performing methods we now concentrate on finding the features that optimise the performance of the model. Again we focus on three classifiers; Logistic Regression, Random Forest and Decision Tree tuned to their optimal parameters (see previous section).  We run 30 models; each model has a number of features ranging from 1 to 30. These features are the best-performing as selected with a forward selection deterministic wrapper described earlier. The results are plotted in Figure \ref{three_classifiers}.
	
	Figure \ref{three_classifiers} suggests that Logistic Regression peaks at a model with seven features and then decays. Similarly, the Random Forest peaks at models with eight and ten features respectively and decays for models with more features. In contrast, the Decision Tree peaks at a model with six features but remains constant (on average) for models with more features. This overall behaviour was also observed in Figure \ref{fig:select_classifier}, where all classifiers, except the Decision Tree, peak their performance at models with four to eight features. Models with more than eight features decrease their accuracy on average.  
	
	For Logistic Regression these seven features are users' followers (user-based), complexity of tweets, defined as the depth of the dependency tree of the tweet (message-based), fraction of tweets denying the rumour (message-based), sentiment of users' past tweets (user-based), tenure of users in Twitter (in days) (user-based), retweets within the network (network-based) and low to high diffusion, defined as a retweet from a user with a higher number of followers than the retweeted user (network-based).
	However, not all of these features are statistically significant for this data set. We further run a statistical test to assess the significance of these features. We use the log-likelihood test and confidence level 0.05~\cite{fox1997applied}. We found that only three out of seven features are statistically significant. These are the fraction of tweets against the rumour, the tenure of users and users' followers. 
		
	Random Forest peaks its performance at a model with eight features, which are the number of propagation trees in the propagation graph (network-based), fraction of tweets denying the rumour (message-based), verified users (user-based), users with location information (user-based), the degree of the root of the largest connected component (network-based), number of users' likes (user-based), quotes within the network (network-based) and tweets with negation (message-based).

	Random Forest is a black-box technique with no clear interpretation of the parameters of its trees. It is a non-parametric machine learning technique hence it is not straight-forward to estimate the statistical significance of its features. Nevertheless, there are techniques that estimate the feature importance. Particularly, the relative rank, i.e. depth, of a feature used in a node of a tree can determine the relative importance of the feature. For example, features used at the top of a tree contribute to the final prediction decision of a larger fraction of the input samples~\cite{hastie2009elements,scikit_learn}. Measuring the expected fraction of the sample they contribute gives an estimate of the relative importance of the features. 
	
	We run the Random Forest model 1,000 times and take the average of the feature importance measure. 
	The relative importance is a number between 0 and 1 and the higher it is the more important a feature is. 
	We find that the fraction of tweets that deny the rumour is the most important feature for classification and the degree of the root of the LCC follows, scoring 0.22 and 0.19 respectively. The features related to the number of trees in the propagation graph and the verified users seem to contribute significantly, 0.16 and 0.15 respectively, while the remaining four play a less important role, scoring less than 0.08.
	
	Figure \ref{three_classifiers} shows that the Decision Tree algorithm peaks its performance at a model with six features. Investigating the resulting Decision Tree and its splitting rules it became evident that actually only three out of six features are used in the decision rules. These are
	(i) the fraction of tweets denying the rumour (message-based), (ii) users with description (user-based) and (iii) user's post frequency (user-based).
	A further analysis unveils that although we feed the Decision Tree with an increasing number of features the algorithm always uses only a small subset of them which never exceeds eight. This justifies why the $F_1$-score of the Decision Tree classifier remains constant on average as we increase the number of features in Figure \ref{fig:select_classifier} and Figure \ref{three_classifiers}. Every time we add a new feature, the Decision Tree uses a small subset with similar performance to the previous models. 
	
	Examining the three models we observe that they have one feature in common, the fraction of users that deny the rumour. Logistic regression and Random Forest have a mixture of message-based, user-based and network-based features, whereas the Decision Tree uses only message-based and user-based features. We strongly believe that with the addition of more rumours and larger data sets the Decision Tree algorithm will use more features among which the network-based ones. In conclusion, a high accuracy can be achieved with a relatively small set of features for this data. This is to be expected as sets with a small number of data points are subject to overfitting when a large number of features are used.
	
	\section{Results}
	\label{results}
	
	In the previous section, we focused on finding the best models among a variety of possibilities. Having determined the best three models, their parameters and their features, we are ready to calibrate and validate the final model. We split the data into training (60\%) and testing (40\%) sets. These two sets also preserve the overall ratio of true to false observations. The results for the test set are presented in Table \ref{test_scores}.
	
	\begin{table}
		\begin{center}
			\begin{tabular}{ l || c| c| c| c| c| c| }
				Test Set & Accuracy & Precision & Recall & $F_1$-score & AUC & kappa \\
				\hline
				Logistic Regression & 0.828  & 0.8 & 0.941 & 0.865 & 0.936 & 0.631 \\
				Random Forest & 0.897 & 0.938 & 0.882 & 0.909 & 0.971 & 0.789  \\
				Decision Tree & \textbf{0.966} & \textbf{1.0} & \textbf{0.941} & \textbf{0.970} & \textbf{0.971} & \textbf{0.930}  \\
			\end{tabular}
			\caption{Classification metrics of the three models for the test set.}
			\label{test_scores}
		\end{center}
	\end{table}
		
	 We highlighted the best-performing model on the test dataset, which is the Decision Tree. It reaches a high accuracy rate, 96.6\% and a precision 1.0. This implies that there are no false-positive predictions, i.e. rumours that are false but classified as true. The recall is 0.94 implying the presence of false negatives.
	
	Random Forest follows with accuracy close to 90\%. Logistic regression achieves the lowest accuracy of the three models, 82.8\%. Although the $F_1$-score of Random Forest model is higher than the $F_1$-score of the logistic Regression their precision and recall scores differ substantially. Random Forest has a higher precision but lower recall than the Logistic Regression. This suggests that the Random Forest model returns a lower number of positives but most of the positives are correctly classified. On the other hand Logistic Regression returns many positives but a higher number of incorrect predicted labels. 
	
	\subsection{Comparison to Benchmark Models}
	\label{compare_to_benchmark}
				
	As discussed previously we are principally interested in determining the veracity of rumours as quickly as possible. In order to test this we split the duration of a rumour into 20 time-intervals and extract all the features from the beginning of the rumour until the end of each time interval. This results in 20 observations per rumour. We apply the three models to each time interval and calculate the veracity of the rumours at each time-step. 
	
\begin{figure}
			\begin{center}
				\includegraphics[scale=0.5]{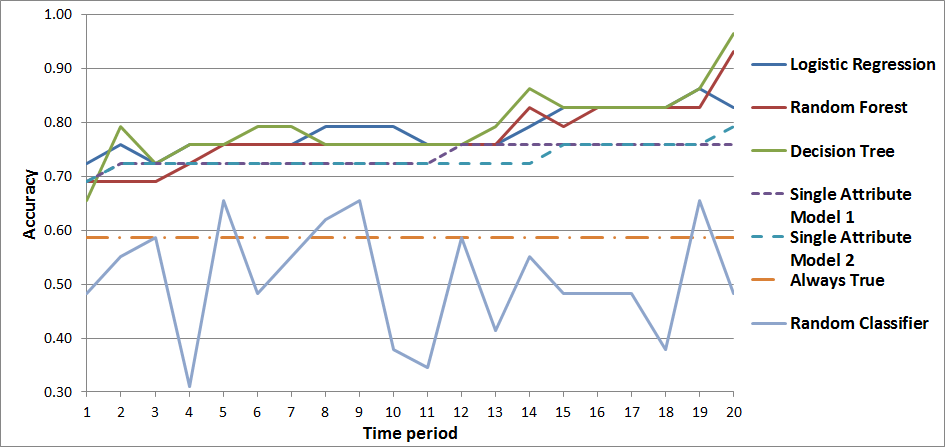}
				\caption{The accuracy as a function of time for different models.}
				\label{benchmark_accur}
			\end{center}
		\end{figure}

	We calculate the accuracy of the three models at each time step. We compare against four benchmark models. The first is a random classifier which randomly classifies a rumour either true or false at each time period. The second model is the ``always true" model, which always predicts that a rumour is true. The third model, named ``single attribute model 1" classifies a rumour as true if the number of tweets supporting at a given time period is greater than the number of tweets against. Otherwise it classifies it as false. The ``single attribute model 2'' is similar to the ``single attribute model 1'', but a rumour is classified true if the ratio of in-favour tweets over the against is greater than 2.22. Otherwise it is false. The number 2.22 is the ratio of total tweets in the dataset that are in favour over the total tweets that are against.  This gives on average how many more supporting tweets exist in the dataset. Figure \ref{benchmark_accur} shows the results. 
	
	We observe that the random classifier oscillates around 0.5 as expected. The two ``single attribute" models are better than both the random classifier and the ``always true" model, with the ``single attribute model 2" performing slightly better than ``single attribute model 1". Our machine learning models perform better than the simple models especially towards the end of the rumour. For example, in the beginning of the rumour the Random Forest and Decision Tree models have similar accuracy as the ``single attribute model 2". However at the end of the rumour the Random Forest and Decision Tree have improved their accuracy by 33\% and 37\% respectively, while the simple model’s accuracy is improved by 14\%. The machine learning models have the ability to improve the accuracy at higher rates than any other simple model. This is a crucial result for decision making at early stages of a rumour development, before an official verification. 
	
	As a result of our analysis we also developed a visualisation tool which summarises and depicts the key observations and results. To avoid taking focus away from the modelling aspects we present further details in Appendix \ref{visualisation_tool}.

	\subsection{Comparison to Literature Findings}
	\label{compare_to_literature}
	
	The relevant papers to our work are summarised in Section \ref{intro}. Here we compare our findings to the results of a few key papers \cite{mendoza2010twitter,castillo2011information,castillo2013predicting,kwon2013prominent,finn2014investigating,vosoughi2015automatic}. Similar to our conclusion, some of these studies \cite{finn2014investigating,mendoza2010twitter} found that the tweets which deny the rumour play a significant role in estimating the veracity of the rumour\footnote{A phenomenon known as the ``wisdom of the crowd''.}.
	However, to the best of our knowledge, the key differences are the following. Firstly, we worked with a larger set of potential features, consisting of 87 features; particularly, none of these studies considered features related to users' past behaviour. Secondly, we aggregated the tweet, user and network attributes to rumour features using a non-trivial formula \eqref{aggregation1} which captures the difference between tweets that support and those that deny the rumour.
	
	Thirdly, we found that we need a lower number of features than other models in the literature \cite{castillo2011information,castillo2013predicting,kwon2013prominent,vosoughi2015automatic}, varying between three and eight. Although the Logistic Regression and Random Forest models admit six and eight features respectively, about three are statistically significant or most important. It is interesting that high accuracy can be achieved with a small number of features. This can be explained by the model overfitting a relatively small number of data points when more features are used. However, we expect that the number of discriminatory features might increase as the volume of data (more data points) and variety of rumours increase. The extra data also allows us to place more emphasis on early classification. This is an open question that we aim to address in the future.
	
	Fourth, the accuracy of our classifiers varies between 82.8\% (Logistic Regression) and 96.6\% (Decision Tree) on ``unobserved data" (validation set). To our best knowledge, the two best models, Random Forest and Decision Tree, outperform any other model in the academic literature \cite{castillo2011information,castillo2013predicting,kwon2013prominent,vosoughi2015automatic}. We achieve a high success rate which shows the potential benefit of our model.
	
	Last and most importantly, we considered the time-evolution of the features and hence the veracity. We built a model which is able to infer the veracity of a rumour with high accuracy at early stages, before the official confirmation from trusted sources. A time-series analysis was first attempted in \cite{castillo2013predicting}, where the authors estimate the rumour veracity before the burst of a rumour (when the activity suddenly increases). Although this approach introduces some dynamical aspects, it lacks a full and detailed time-series analysis. 
	Later, a proper time-series analysis of veracity was performed in \cite{vosoughi2015automatic}. The author introduces two classifiers which are specifically designed for time-series classification, the Dynamical Time Wrapping (DTW) and Hidden Markov Model (HMM) classifiers. 
	
	These two models achieve an accuracy of 71\% and 75\% respectively using 37 features. From these features only 17 were found to be statistically significant.  The author modelled the veracity of rumours at different time-periods (as percentage of the time elapsed for each rumour). His best model does not exceed an accuracy rate 60\% at a time-period half-way from the start until the trusted verification of the rumour. In contrast, we achieve a higher accuracy, at least 76\%, at the same time-period, (time-period 10 in Figure \ref{benchmark_accur}). 
	This time-period, on average, corresponds to 3 hours and 20 minutes after the outbreak of the rumour. A 76\% accuracy is already reached by all of our models at one quarter of the rumour duration, which on average corresponds to 1 hour and 50 minutes after the beginning of the rumour. However, as the time passes and more tweets and information are obtained, understandably our model accuracy increases. With more modelling time and more data, we would hope to improve early declaration still further.
	
	\section{Conclusion and Future Work}
	\label{discussion}
	
	Modern lifestyle heavily relies on social media interaction and spread of information. New challenges have emerged as large volumes of information are being propagated across the internet. Assessing the trustworthiness of the news and rumours circulating in a network is the main subject of this study. The primary goal of this paper is to develop the core algorithms that can be used to automatically assign a trustworthiness measure to any communication. 
	
	We collected 72 rumours and extracted 87 features which capture three main topics and derived from our reflection on the relevant literature. The topics are the linguistic characteristics of the messages, the users' present and past behaviour and how the information propagates through the network.  
	Furthermore, the feature space encompasses dynamic aspects for estimating rumour veracity, contributing to the literature since only one study thus far has attempted a similar approach. In addition to the modelling, we developed a graphical user interface which allows the user to investigate in details the rumour development over time. 
	
	Our overall model was significantly more accurate than similar studies due to two main reasons: (i) introduction of novel features, e.g. users' past behaviour, and (ii) the  method of aggregating tweet/user attributes to rumour features.  
	Our key findings suggest that the  Decision Tree, Random Forest and Logistic Regression are the best classifiers. Additionally, the fraction of tweets that deny the rumour plays an essential role in all models. Finally, the three models require only a low number of features, varying between three and eight. 
	
	Although our paper provides the first and decisive step towards a system for determining the veracity of a rumour, there are opportunities for further research which will enhance our system. The automation of the rumour collection and tweet annotation is one area for future work. In our system the categorisation of the tweets into rumours is a manual and time-consuming task. Similarly, the annotation of the tweets require much effort and time from our side. For these reasons, we aim to build another classifier that automatically classifies the tweets into rumours and annotates them based on the content of text. This way we will be able to collect a large volume of rumours and massively scale up our dataset. Having a larger volume of data and more diverse rumours will allow us to develop more robust and accurate models.
	
	The current models return either the probability of a rumour being true or the class itself. There is no information about the confidence levels of the results. One of the main future goals is to produce an algorithm providing uncertainty estimates of the veracity assessments. Additionally, we would like to expand our data sources and consider data from other social networks, such as the YouTube platform. Calibrating and testing our model on other sources of data will give further confidence about its validity and will extend its applicability.

	\section*{Acknowledgements}
	This work was partly supported by UK Defence Science and Technology Labs under Centre for Defence Enterprise grant CDE42008. We thank Andrew Middleton for his helpful comments during the project. We would also like to thank Nathaniel Charlton and Matthew Edgington for their assistance in collecting and preprocessing part of the data.

	\section*{Appendix}
	\label{appendix}
	\appendix
		
	\section{Data Collection Process}
	\label{data_collection}
	Our data collection process consists of four main steps:
	\begin{enumerate}
		\item Collection of tweets with a specific keyword, e.g. ``\#ParisAttacks" or ``Brussels". The Twitter API only allows the collection of such tweets within a ten-day window. For this reason this step must start as soon as an event happens or a rumour begins.  
		\begin{enumerate}
			\item Manual analysis of tweets and search for rumours. In this step we filter out all the irrelevant tweets. For example, if we collected tweets containing the keyword ``Brussels" (due to the unfortunate Brussels attacks), we ignore tweets talking about holidays in Brussels.
			\item Collection of more tweets relevant to the story with keywords that we missed in the beginning of Step 1 (this step is optional). For example, while searching for rumours we might come across tweets talking about another rumour. We add the keyword that describes this new rumour in our tweet collection. 
			\item Categorise tweets into rumours. Group all tweets referring to the same rumour. 
			\item Identify all the unique users involved in a rumour. This set of users will be used in Steps 2 to 4.
		\end{enumerate}
		\item Collect users' most recent 400 tweets, posted before the start of the rumour. This step is required because we aim to examine the users' past behaviour and sentiment, e.g. whether users' writing style or sentiment changes during the rumour, and whether these features are significant for the model. To the best of our knowledge, this set of features is considered for the first time in the academic literature in building a rumour classifier.
		\item Collect users' followees (friends). This data is essential for making the propagation graph, see Section \ref{data} and Appendix \ref{making_prop_graph}.
		\item Collect users' information, including user's registration date and time, description, whether account is verified or not etc.
	\end{enumerate}
	
	\subsection{Rumours Summary Statistics}
	\label{summary_stats}
	We provide a summary statistics table of the 72 collected rumours, see Table \ref{table:tweetsummary}. This table shows the total number, mean, median, etc., of the distributions of the number of tweets, the percentage of supporting tweets, etc., of the 72 rumours, as well as some statistics of four example rumours. 
	We collected about a 100 million tweets, including users' past tweets. From the collected tweets, about 327.5 thousand tweets are part of rumours.  These tweets contributed to the message-based features of the classification methods. The users' past tweets contributed only to the features capturing a user's past behaviour.  
	\begin{table}
		\begin{center}
			\begin{tabular}{ | l || r | r | r | r | r | r | }
			\hline
						& Tweets 	& \%Support & \%Against & Users & Users Tweets 	& Duration(hours)   \\ \hline \hline
			Total 		& 327,484	& 60.9\% 	& 27.4\% & 270,054 	& 95,579,214	& N/A 	\\ \hline
			Mean 		& 4,548  	& 65.7\% 	& 22.9\% & 3,751 	& 1,327,489 	& 9.02 	\\ \hline
			Median 		& 1,660  	& 81.5\% 	& 2.5\%  & 1,540 	& 520,288 		& 3.04 	\\ \hline
			Std 		& 6,816  	& 34.4\% 	& 32.2\% & 5,146 	& 2,005,616 	& 16.40 \\ \hline
			Min 		& 23 	 	& 0.3\% 	& 0.0\%  & 23 		& 9,553 		& 0.07 	\\ \hline
			Max 		& 46,807 	& 100.0\% 	& 97.5\% & 32,529 	& 13,877,121 	& 114.22 \\ \hline
			Example 1 	& 46,807 	& 76.1\%	& 1.3\%  & 32,529 	& 13,877,121 	& 14.42	\\ \hline
			Example 2 	& 18,525 	& 82.2\%	& 9.6\%	 & 16,081	& 5,852,204		& 1.37 	\\ \hline
			Example 3 	& 71		& 53.5\%	& 8.5\%	 & 69		& 24,303		& 3.84 \\ \hline
			Example 4 	& 23		& 26.1\%	& 43.5\% & 23		& 9,553			& 3.65 \\ \hline 
			\end{tabular}
			\caption{A summary statistics of the collected rumours. Examples 1 and 2 correspond to the rumours with the largest and second largest number of tweets respectively. Examples 3 and 4 correspond to the rumours with the second smallest and smallest number of tweets respectively.}
			\label{table:tweetsummary}
		\end{center}
	\end{table}

	\section{Making the Propagation Graph}
	\label{making_prop_graph}
	Nodes in the propagation tree correspond to unique users. Edges are drawn between users who retweet messages. However the retweet relationship cannot be directly inferred from the Twitter data. Consider a scenario with three users, A, B and C. User A posts an original tweet. User B sees the tweet from user A and retweets it. Twitter API returns an edge between user A and user B. If user C sees the tweet from user B and retweets it, Twitter API returns an edge between the original user A and user C, even though user A is not a friend with user C and there is no way user C could have seen the tweet from user A. To overcome this, we have collected the users’ followees. Therefore, in our scenario user B is connected to user C only if the retweet timestamp of user C is later than the retweet of user B and user B is in the followees’ list of user C.
	
	\section{A practical example for using formula \eqref{aggregation1}}
	\label{practexample}
	Here, we elaborate on formula \eqref{aggregation1} and present a practical example. For simplicity reasons and to avoid confusion we define support, $S^{(i)}$, neutral, $N^{(i)}$, and against, $A^{(i)}$, terms in formula \eqref{aggregation1} following the example attributes given in Section \ref{feats}. The generalisations are straightforward.  If the attribute of the tweet is a binary indicator, for example whether a tweet contains a URL link or not, we define
	\begin{eqnarray}
		S^{(i)} &=& \frac{\text{number of tweets with url that support the rumour}}{\text{total number of tweets that support the rumour}}, \nonumber\\
		N^{(i)} &=& \frac{\text{number of tweets with url that are neutral to the rumour}}{\text{total number of tweets that are neutral to the rumour}}, \nonumber\\
		A^{(i)} &=& \frac{\text{number of tweets with url that deny the rumour}}{\text{total number of tweets that deny the rumour}}. \nonumber
	\end{eqnarray}

	If the attribute of the tweet is continuous, for example, the number of words in a tweet, we then define 
	\begin{eqnarray}
		S^{(i)} &=& \text{average number of words in tweets that support the rumour}, \nonumber\\
		N^{(i)} &=& \text{average number of words in tweets that are neutral to the rumour}, \nonumber\\
		A^{(i)} &=& \text{average number of words in tweets that deny the rumour}. \nonumber
	\end{eqnarray}
	
	These expressions are then combined through formula \eqref{aggregation1} to give the relevant feature of the rumour.

	\section{Feature Reduction Methods}
	\label{feat_reduction}
	
	Since our dataset consists of 72 rumours, from theoretical and experimental arguments, we expect the relevant features to be about 10. We expect models with as many as 20 features to begin to show a decrease in performance. For this reason we set the upper bound on the number of features to be 30 and aim to examine models with an increasing number of features from 1 to 30. If this bound proves to be low we will reconsider this choice. However as it becomes evident in Section \ref{analysis}, this bound is satisfactory.
		
	In this study we use four methods which are combinations of those described so far. For filtering we use the ANOVA F-test \cite{lomax2012introduction}. 
	
	\begin{enumerate}[label=Method \arabic*.]
		\item A combination of filter method, random wrapper and deterministic wrapper
		\begin{enumerate}
			\item Use ANOVA F-Statistics for filtering. Keep the 30-best scoring features. 
			\item From those 30-best we applied the classifier to 100,000 different combinations of 3 features to find the combination of 3 which maximise the $F_1$-score.
			\item Add one-by-one the remaining 27 features by applying the classifier and keeping the one with the best $F_1$-score in each round. 
		\end{enumerate}
		\item A forward selection deterministic wrapper method
		\begin{enumerate}
			\item Apply the classifier to all features individually and select the one which maximises the $F_1$-score (from all available features, no pre-filtering is required). 
			\item Scan (by applying the classifier) all remaining features to find the combination of two (one from step a.) that maximises the $F_1$-score.
			\item Continue adding one-by-one the features which maximise the $F_1$-score until the number of features reaches 30.
		\end{enumerate}
		\item A combination of filter method and forward selection method
		\begin{enumerate}
			\item Use the ANOVA F-Statistics for filtering and keep the 30-best scoring features. 
			\item Apply the classifier and find the best-scoring, i.e. maximum $F_1$-score, from the 30-best selected from the filtering method (step a.).
			\item Continue adding one-by-one the features which maximise the classification $F_1$-score.
		\end{enumerate}
		\item A feature transformation method
		\begin{enumerate}
			\item Use a feature transformation method, the principal component analysis. Keep the 30-best components. 
			\item Start with the principal component from the 30-best selected from step a.
			\item Start adding the components one after the other.
		\end{enumerate}
	\end{enumerate}

\begin{figure}[t]
		\begin{center}
			\includegraphics[scale=0.5]{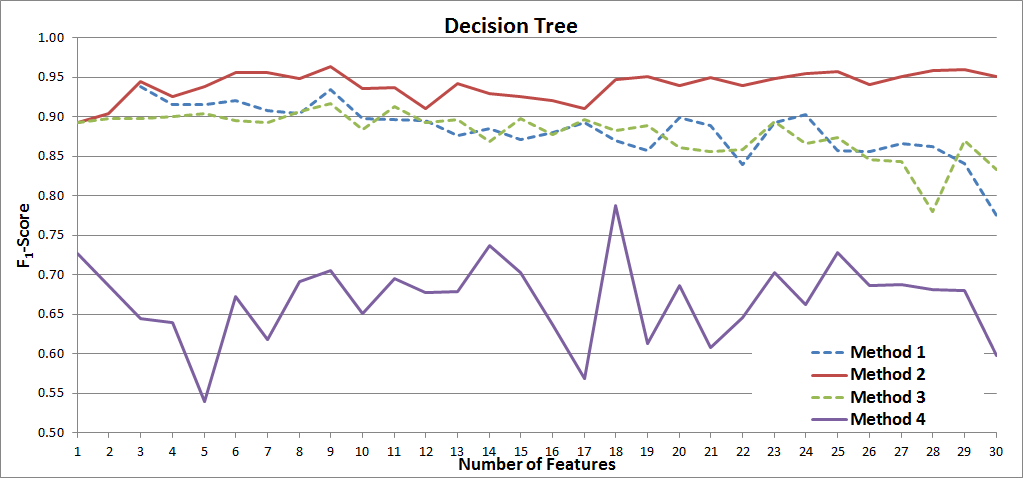}
			\caption{$F_1$-score for Decision Tree versus the number of features/components selected from four methods of feature reduction.}
			\label{feat_reduction_methods_RF}
		\end{center}
	\end{figure}

	We apply each method to each classifier separately, using \textit{scikit-learn}'s default parameters, and assess it using k-fold cross validation. We have abandoned the Neural Network method for two reasons. First its performance was poor compared to the other methods and secondly it required long computational times which slowed down considerably the analysis of the results.  We plot the $F_1$-score as a function of the number of features for the remaining classifiers and each feature reduction method, see Figure \ref{feat_reduction_methods_RF}.
		
	We observe that the second method (red line in Figure \ref{feat_reduction_methods_RF}) outperforms, in almost all cases, all the other techniques. Similar plots are produced and same conclusion is reached for the other classifiers too. Therefore we can safely conclude that the forward selection deterministic wrapper is consistently the best-performing method of feature reduction for all classifiers. 
		
	\section{Further Results on Classifier Selection}
	In Section \ref{analysis} we present the results from running several classifiers for thirty models, each model having an increasing number of features from one to thirty.  Here we present more results that support our choice for feature selection. 
	
	\begin{figure}[t]
		\begin{center}
			\includegraphics[scale=0.5]{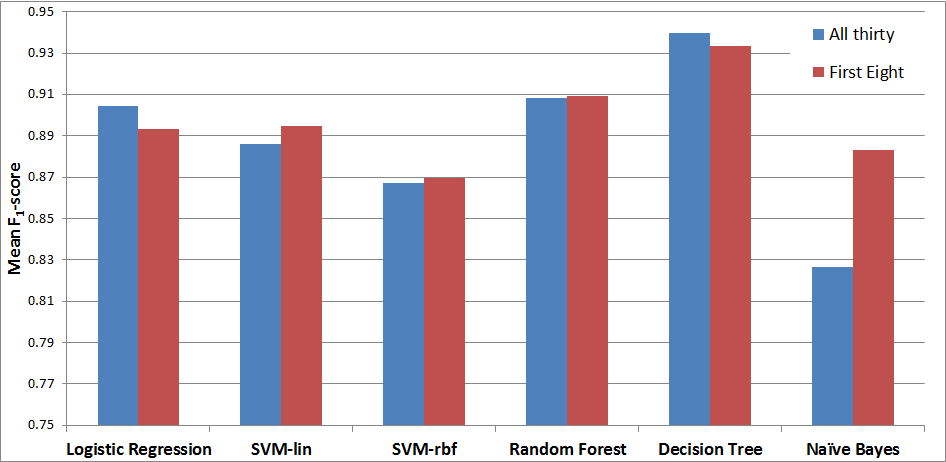}
			\caption{Mean $F_1$-score of 30 (blue) and first 8 (red) models.}
			\label{select_classifier2}
		\end{center}
	\end{figure}
	
	In Figure \ref{select_classifier2} we plot the average $F_1$-score for each method. This is a two-column plot. The first column (blue) corresponds to the average $F_1$-score of all 30 models.  The second column (red) is the average $F_1$-score of the first eight models (those with number of features from 1 to 8)\footnote{We compute the average of the first eight models because this is the range where the classifiers peak their performance. As we argue in Section \ref{analysis} all plots indicate that classifiers’ performance decreases when more than eight features are added.}. 
	
	\label{select_classifier}

	\section{Visualisation Tool}
	\label{visualisation_tool}
\begin{figure}
		\begin{center}
			\includegraphics[scale=0.27]{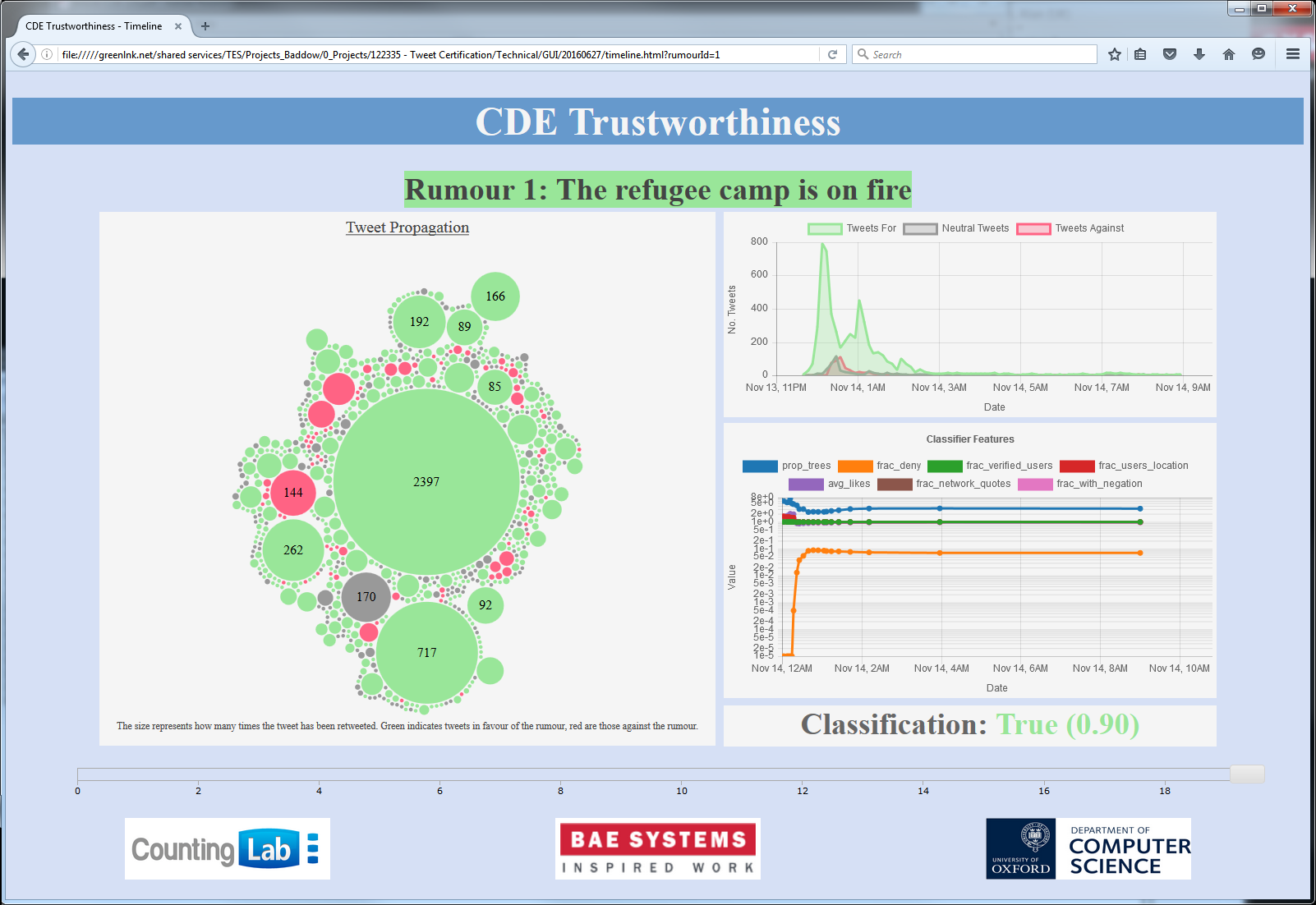}
			\caption{The visualisation tool. Inside a rumour.}
			\label{fig:visualisation_tool}
		\end{center}
	\end{figure}

	As a by-product of our modelling, we also developed a software tool which helps the user to visualise the results and gain a deeper understanding of the rumours, see Figure \ref{fig:visualisation_tool}. The tool consists of three layers. On the first layer the user selects a topic of interest (e.g. ``Paris Attacks''). This directs to the second layer which displays all the relevant rumours with a basic summary (e.g. the rumour claim, timestamp of the first tweet, a word cloud, distribution of the tweets that are in favour, neutral or against the rumour and the modelled veracity). After selecting a rumour of interest, the user is navigated to the third layer, shown in Figure \ref{fig:visualisation_tool}. There, the tool shows several figures, such as the propagation forest (supporting, neutral and denying trees are coloured in green, grey and red respectively), a histogram showing the number of tweets in favour of the rumour, against the rumour, and those that are neutral, a plot of classifier's features and the rumour veracity. A time-slider is provided to allow the user to navigate through the history of the rumour by selecting one of the available time steps. Moving the slider the user can investigate how the rumour, its veracity and the key features evolve over time. This gives the flexibility to the user to explore the key factors that affect the veracity of the rumour.
	
	\bibliography{bibliography}
	\bibliographystyle{plain}
	
\end{document}